\documentclass[twocolumn,prb,aps,showpacs,amsmath,amssymb]{revtex4}

\usepackage{graphicx}
\usepackage{dcolumn}
\usepackage{bm}

\begin{document}

\title{Keldysh effective action theory for universal physics in spin-$\frac{1}{2}$ Kondo dots}

\author{Sergey Smirnov}
\email{sergey.smirnov@physik.uni-regensburg.de}
\author{Milena Grifoni}
\affiliation{Institut f\"ur Theoretische Physik, Universit\"at Regensburg,
  D-93040 Regensburg, Germany}

\date{\today}

\begin{abstract}
We present a theory for the Kondo spin-$\frac{1}{2}$ effect in strongly
correlated quantum dots. The theory is applicable at any temperature and
voltage. It is based on a quadratic Keldysh effective action parameterized by
a universal function. We provide a general analytical form for the tunneling
density of states through this universal function for which we propose a
simple microscopic model. We apply our theory to the highly asymmetric
Anderson model with $U=\infty$ and describe its strong coupling limit, weak
coupling limit and crossover region within a single analytical expression. We
compare our results with numerical renormalization group in equilibrium and
with a real-time renormalization group out of equilibrium and show that the
universal shapes of the linear and differential conductance obtained in our
theory and in these theories are very close to each other in a wide range of
temperatures and voltages. In particular, as in the real-time renormalization
group, we predict that at the Kondo voltage the differential conductance is
equal to $2/3$ of its maximum.
\end{abstract}

\pacs{72.15.Qm, 73.63.-b, 72.10.Fk}

\maketitle

The Kondo state of a quantum dot (QD) continues to attract significant
interest from both experiment and theory. The zero-dimensional platform
available in modern nanotechnology provides a great advantage over bulk
metals with magnetic impurities where the Kondo effect was originally
observed \cite{de_Haas_1934}. Indeed, the possibility to enrich the physics
with nonequilibrium as well as to independently tune various parameters of
QD structures allows one to access different aspects of the Kondo physics
predicted theoretically. One of these aspects is the universality of the
zero bias anomaly or Kondo resonance in the differential conductance which,
as has also been confirmed in experiments \cite{Goldhaber-Gordon_1998a,
Grobis_2008}, turns out to be a universal function of the temperature and
voltage with the energy scale given by the Kondo temperature $T_\text{K}$.
Another advantage of the QD framework is the possibility to study the fate
of the many-particle Kondo resonance in the presence of external fields or
when the QD is coupled to contacts with nontrivial ground states. Recent
experiments have demonstrated the universality of the Kondo effect also
within these advanced setups, {\it e.g}., with ferromagnetic contacts
\cite{Gaass_2011} and external magnetic fields \cite{Kretinin_2011}. Other
aspects of the Kondo state, such as its use for spin manipulations
\cite{Hauptmann_2008} or ferromagnetic-superconducting correlations
\cite{Hofstetter}, have been addressed in modern experiments.

The whole diversity of the Kondo physics in QDs can be well captured within
the single-impurity Anderson model (SIAM) \cite{Anderson_1961}. Many
theoretical concepts have been developed to solve SIAM for QDs in the Kondo
state in equilibrium and nonequilibrium. All these theories can be in
general classified with respect to their applicability to two possible
regimes of the Kondo state: the weak coupling regime, when the temperature
is higher than the Kondo temperature, $T>T_\text{K}$, and the strong
coupling regime, when $T<T_\text{K}$. Among the theories which can access
only the weak coupling regime are, {\it e.g.}, semi-analytical perturbation
theories \cite{Sivan_1996} well above $T_\text{K}$, or advanced analytical
slave-bosonic Keldysh field integral theories
\cite{Smirnov_2011,Smirnov_2011a} which extend the applicability range
close to $T_\text{K}$. In the strong coupling regime for temperatures
$T\ll T_\text{K}$ there are slave-bosonic numerical mean field
theories \cite{Aguado_2000,Lopez_2003}. From high temperatures to
temperatures not too much below $T_\text{K}$ the noncrossing
approximation (NCA) \cite{Wingreen_1994,Haule_2001} is a numerical tool
which, however, predicts a wrong scaling, $T_\text{NCA}\neq T_\text{K}$. A
less quantitatively reliable theory, based on the method of equations of
motion, may be used for a qualitative description in the whole temperature
range. It has been applied, {\it e.g.}, to QDs in an external magnetic
field \cite{Meir_1993}. The whole temperature range can be comprehensively
numerically described in equilibrium by the numerical renormalization
group (NRG) method \cite{Bulla_2008} which is quite flexible with respect
to different physical setups and has been used, {\it e.g.}, for QDs coupled
to ferromagnetic contacts \cite{Sindel_2007}. In nonequilibrium the
recently developed semi-analytical real-time renormalization group (RTRG)
method \cite{Pletyukhov_2012} is a promising theory.

The whole spectrum of modern theoretical methods is in principle enough to
describe many fundamental properties of the Kondo state in
spin-$\frac{1}{2}$ QDs in different regimes. However, a simple single
analytical theory which could provide a reliable quantitative description of
the Kondo physics at any temperature and voltage and which would elucidate its
universality is highly desirable.

Here we make a fundamental step towards such a theory using the Keldysh field
integral formulation in the slave-bosonic representation. Namely, we provide
the tunneling density of states (TDOS) in the Kondo regime, Eqs. (\ref{tdos})
and (\ref{rse}), and reduce the problem to finding a universal function of the
temperature and voltage which enters this general form of the TDOS.

We apply our theory to the highly asymmetric spin-$\frac{1}{2}$ SIAM with a
strong electron-electron interaction, $U=\infty$ (extension to finite $U$ is
straightforward), and finite $\mu_0-\epsilon_\text{d}$ ($\epsilon_\text{d}$ is
the single-particle QD energy level, $\mu_0$ is the equilibrium chemical
potential) at any temperature and voltage. The knowledge on this model up to
now has been restricted by slave-bosonic mean field theories
\cite{Aguado_2000,Lopez_2003} deep below $T_\text{K}$ or, above $T_\text{K}$, by
NCA \cite{Wingreen_1994,Haule_2001} and the Keldysh field theory
\cite{Smirnov_2011}.

An exact description of the universal linear conductance can be accessed within
NRG. However, little attention has been paid to its prediction for the linear
conductance in the highly asymmetric SIAM. But what is more important is that
NRG fails out of equilibrium and the only nonequilibrium theory able to cover
the whole voltage range, RTRG \cite{Pletyukhov_2012}, is applicable only to the
$s-d$ model. Thus, out of equilibrium the situation with the highly asymmetric
SIAM with $U=\infty$ is even worse.

Within the present work we eliminate this lack of knowledge on the Kondo
regime of the highly asymmetric spin-$\frac{1}{2}$ SIAM with $U=\infty$ by
deriving an analytical expression for the TDOS which provides 1) Linear and
differential conductance in the whole range of temperatures and voltages with
the correct scaling $T_\text{K}$; 2) Fermi liquid theory (quadratic
temperature and voltage dependence of the differential conductance) at low
energies; 3) Analytical ratio between the corresponding Fermi liquid
coefficients. This ratio has been intensively discussed in the literature
\cite{Oguri_2005,Sela_2009} for the symmetric spin-$\frac{1}{2}$ SIAM. However,
little is known on this ratio for the highly asymmetric spin-$\frac{1}{2}$ SIAM
with $U=\infty$; 4) Prediction identical to the one made in the RTRG
\cite{Pletyukhov_2012} for the $s-d$ model that in the deep nonequilibrium
crossover region, at the Kondo voltage, the differential conductance is equal
to $2/3$ of its maximum; 5) Excellent agreement with the Hamann's analytical
theory \cite{Hamann_1967} for the $s-d$ model at high temperatures.

Let us briefly recall the slave-bosonic approach used in Ref.
\onlinecite{Smirnov_2011}. The original QD spin-$\frac{1}{2}$ fermionic operators of
the Anderson Hamiltonian, $d_\sigma$, $d_\sigma^\dagger$ ($\sigma=\uparrow,
\downarrow$), are expressed in terms of new spin-$\frac{1}{2}$ fermionic,
$f_\sigma$, $f_\sigma^\dagger$, and slave-bosonic, $b$, $b^\dagger$, operators. Since
the double occupancy is forbidden, these operators satisfy the constraint,
$\hat{Q}=\hat{I}$ with $\hat{Q}\equiv b^\dagger b+\sum_\sigma f_\sigma^\dagger f_\sigma$.
In practical calculations of observables this restriction is taken into account in
the Keldysh field integral \cite{Altland_2010} representation of a given QD
observable $\hat{O}=\mathcal{F}(d_\sigma^\dagger$, $d_\sigma)$ via the replacement
of the QD Hamiltonian $\hat{H}_\text{QD}$ with $\hat{H}_\text{QD}+\mu\hat{Q}$,
where $\mu$ is a positive real parameter with respect to which one takes the
limit $\mu\rightarrow\infty$ at the end of the calculation,
\begin{equation}
\begin{split}
&\langle\hat{O}\rangle(t)=\frac{1}{\mathcal{N}_0}\underset{\mu\rightarrow\infty}{\text{lim}}e^{\beta\mu}
\int\mathcal{D}[\bar{\chi},\chi]e^{\frac{i}{\hbar}S_\text{eff}[\bar{\chi}^\text{cl,q}(\tilde{t});\chi^\text{cl,q}(\tilde{t})]}\\
&\times\mathcal{F}[\bar{\chi}^\text{cl}(t),\bar{\chi}^\text{q}(t);\chi^\text{cl}(t),\chi^\text{q}(t)].
\end{split}
\label{KFI_observ}
\end{equation}
Here $\chi^\text{cl}(t)$ and $\chi^\text{q}(t)$ (and $\bar{\chi}^\text{cl}(t)$,
$\bar{\chi}^\text{q}(t))$ are the classical and quantum components (and their
conjugate partners) of the QD slave-bosonic field which is the coherent state
field of the slave-boson annihilation operator $b$, $S_\text{eff}$ is the Keldysh
effective action which depends on $\mu$ and governs the dynamics of the QD
coupled to the contacts, $\beta\equiv 1/kT$ is the inverse temperature and
$\mathcal{N}_0$ is the normalization constant given in Ref. \onlinecite{Smirnov_2011}.

The Keldysh effective action is a nonlinear functional of the slave-bosonic
fields. It represents the sum of the standard quadratic free slave-bosonic
action $S_0[\bar{\chi}^\text{cl}(t),\bar{\chi}^\text{q}(t);\chi^\text{cl}(t),\chi^\text{q}(t)]$
and the complex nonlinear tunneling action
$S_\text{T}[\bar{\chi}^\text{cl}(t),\bar{\chi}^\text{q}(t);\chi^\text{cl}(t),\chi^\text{q}(t)]=-i\hbar\,\text{tr}\ln\bigl[-iG^{(0)-1}-i\mathcal{T}\bigl]$,
where the matrix $G^{(0)}$ and the matrix $\mathcal{T}$ are defined
in Ref. \onlinecite{Smirnov_2011}: the matrix $G^{(0)}$ describes the isolated QD and
contacts while $\mathcal{T}$ accounts for the QD-contacts tunneling coupling.
The dependence of the tunneling action on the slave-bosonic fields
$\chi^\text{cl}(t)$ and $\chi^\text{q}(t)$ comes through the matrix
$\mathcal{T}$.

We now perform a time-independent shift of the classical components of the
slave-bosonic field in the matrix $\mathcal{T}$,
$\chi^\text{cl}(t)\rightarrow\chi^\text{cl}(t)-\delta\sqrt{2}$,
$\bar{\chi}^\text{cl}(t)\rightarrow\bar{\chi}^\text{cl}(t)-\gamma\sqrt{2}$,
while leaving the quantum components $\chi^\text{q}(t)$, $\bar{\chi}^\text{q}(t)$
unchanged. Note that in general $\gamma\neq\bar{\delta}$ since $\chi^\text{cl}(t)$
and $\bar{\chi}^\text{cl}(t)$ are independent integration variables. This results in
the appearance of nondiagonal blocks in the matrix $G^{(0)-1}$ which now takes the
form:
\begin{equation}
G^{(0)-1}=
\begin{pmatrix}
G^{(0)-1}_\text{d}(\sigma t|\sigma't')&M_\text{T}^{(0)\dagger}(\sigma t|a't')\\
M_\text{T}^{(0)}(at|\sigma't')&G^{(0)-1}_\text{C}(at|a't')
\end{pmatrix},
\label{G0_matr}
\end{equation}
where the inverse of the diagonal blocks, $G^{(0)}_\text{d,C}$, are the same as
in Ref. \onlinecite{Smirnov_2011} and the nondiagonal blocks are given as follows
(for $M_\text{T}^{(0)\dagger}$ $\gamma$ is replaced with $\delta$),
\begin{equation}
M_\text{T}^{(0)}(at|\sigma t')=\frac{1}{\hbar}\delta(t-t')\gamma T_{a\sigma}
\begin{pmatrix}
1&0\\
0&1
\end{pmatrix},
\label{tun_matr_0}
\end{equation}
where $a$ is the set of quantum numbers describing the contacts states  and
$T_{a\sigma}$ are the matrix elements of the tunneling Hamiltonian of
SIAM. It is further assumed that $a=\{q,\sigma\}$, where $q$ describes the
contacts orbital degrees of freedom, $\sigma$ is the contacts spin and
$T_{q\sigma'\sigma}=\delta_{\sigma'\sigma}\tau$. The matrix $\mathcal{T}$
remains, as in Ref. \onlinecite{Smirnov_2011}, off-diagonal but the off-diagonal
blocks change (for $M_\text{T}^\dagger$ $\gamma$ is replaced with $\delta$),
\begin{equation}
\begin{split}
&M_\text{T}=\frac{\delta(t-t')T_{a\sigma}}{\sqrt{2}\hbar}
\begin{pmatrix}
\bar{\chi}^\text{cl}(t)-\gamma\sqrt{2}&\bar{\chi}^\text{q}(t)\\
\bar{\chi}^\text{q}(t)&\bar{\chi}^\text{cl}(t)-\gamma\sqrt{2}
\end{pmatrix}.
\end{split}
\label{tun_matr}
\end{equation}

Up to this point the theory is formally exact but obviously cannot be
solved. To make it tractable we expand the tunneling action
$S_\text{T}$. The crucial difference with respect to Ref. \onlinecite{Smirnov_2011}
is that now we do not use the second order expansion in the variables
$\chi^\text{cl,q}(t)$ (and their conjugates) but instead
we expand up to the second order in the variables
$\chi^\text{cl}(t)-\delta\sqrt{2}$, $\bar{\chi}^\text{cl}(t)-\gamma\sqrt{2}$
and $\chi^\text{q}(t)$, $\bar{\chi}^\text{q}(t)$.

At first sight the second order expansion of the tunneling action in the
variables $\chi^\text{cl}(t)-\delta\sqrt{2}$,
$\bar{\chi}^\text{cl}(t)-\gamma\sqrt{2}$ and $\chi^\text{q}(t)$,
$\bar{\chi}^\text{q}(t)$ may appear inadequate with respect to the Kondo
physics. Indeed, this expansion contains linear terms in the original
slave-bosonic fields $\chi^\text{cl}(t)$, $\bar{\chi}^\text{cl}(t)$ and
$\chi^\text{q}(t)$, $\bar{\chi}^\text{q}(t)$. These terms shift the minimum of
the effective action from the zero slave-bosonic field configuration to a
finite one. Thus the symmetry is broken and the effective action of this type
cannot describe the Kondo state having, as is well-known, no symmetry
breaking.

However, the reasoning above misses one important aspect related to the
Hilbert space on which the Keldysh effective action is defined. The point is
that at fixed $\mu$ this Hilbert space is much wider than the QD Fock space
having states only with zero and one electron. It is easy to show that when
the Hilbert space is narrowed to the physical Fock space of the QD, {\it i.e.},
when the limit $\mu\rightarrow\infty$ is taken, the linear terms in the Keldysh
effective action do not generate any finite contribution to the physical
observables of the QD. Thus they can be discarded from the outset and the
symmetry of the Keldysh effective action is restored.

In this way one obtains a Keldysh effective action which is quadratic with
respect to the original slave-bosonic variables $\chi^\text{cl}(t)$,
$\chi^\text{q}(t)$ (and their conjugates) and where the shifts $\delta$ and
$\gamma$ have been absorbed into the kernel of the action. After this
transformation the theory becomes formally identical to the one in
Ref. \onlinecite{Smirnov_2011}. The only difference is that now the self-energies in
the quadratic action are parameterized by the shifts $\delta$ and $\gamma$.

Therefore, one can immediately write down the general form for the QD TDOS in
the Kondo regime,
\begin{equation}
\nu_\sigma(\epsilon)=(\Gamma/2\pi)\{[\epsilon_\text{d}-\epsilon+\Gamma\Sigma_\text{R}(\epsilon)]^2+[\Gamma\Sigma_\text{I}(\epsilon)]^2\}^{-1},
\label{tdos}
\end{equation}
where $\Gamma\equiv 2\pi\nu_\text{C}|\tau|^2$ ($\nu_\text{C}$ is the contacts
density of states).
\begin{figure}
\includegraphics[width=7.7 cm]{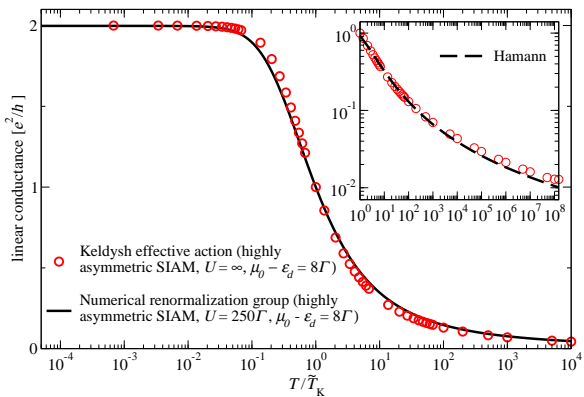}
\caption{\label{figure_1} (Color online) The quadratic slave-bosonic Keldysh
field integral theory (circles) provides the universal equilibrium description
of the Kondo spin-$\frac{1}{2}$ effect in the asymmetric SIAM with $U=\infty$
at any temperature, {\it i.e.}, in the weak coupling regime, strong coupling
regime and in the crossover region. The NRG results (solid line) are provided
by T. Costi and L. Merker (research center of J\"ulich). The inset compares our
theory with the one of Hamann \cite{Hamann_1967}, which is known to describe
well the high temperature regime. In this plot $\widetilde{T}_\text{K}$ is
defined as the temperature at which the differential conductance maximum reaches
half of its unitary limit value. One gets $\widetilde{T}_\text{K}\approx 1.47\,T_\text{K}$.}
\end{figure}

The crucial change with respect to Ref. \onlinecite{Smirnov_2011} concerns the
retarded self-energy
$\Sigma^+(\epsilon)=\Sigma_\text{R}(\epsilon)+i\Sigma_\text{I}(\epsilon)$
which now has a nontrivial dependence on $\alpha\equiv\delta\gamma$,
\begin{equation}
\begin{split}
&\Sigma^+(\epsilon)=\frac{1}{2\pi}\sum_x\biggl[\text{Re}\,\psi\biggl(\frac{1}{2}+\frac{W}{2\pi kT}\biggl)-\\
&-\psi\biggl(\frac{1}{2}+\frac{E_\alpha}{2\pi kT}-\frac{i\mu_x}{2\pi kT}+\frac{i\epsilon}{2\pi kT}\biggl)+i\frac{\pi}{2}\biggl],
\end{split}
\label{rse}
\end{equation}
where the sum is over the contacts ($x=\text{L,R}$), $\psi(z)$ the digamma function,
$\mu_\text{L,R}\equiv\mu_0\mp eV/2$ ($V$ is the bias voltage), $W$ is the half-width
of the contacts Lorentzian density of states ($W\gg kT,eV,\mu_0$) and
$E_\alpha\equiv\alpha\Gamma/2$ is a complex energy.
\begin{figure}
\includegraphics[width=7.7 cm]{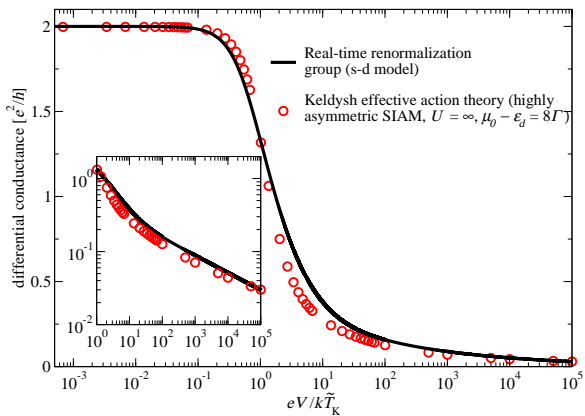}
\caption{\label{figure_2} The universal differential conductance as a function
of the bias voltage at $T=0$ in the Keldysh field integral theory for the
asymmetric SIAM with $U=\infty$ (circles) and in the RTRG theory for the $s-d$
model (solid line). The RTRG data are provided by M. Pletyukhov and H. Schoeller
(University of Aachen). The inset shows the high voltage asymptotics of these two
models. Although the asymmetric SIAM and $s-d$ model are different, their
universal shapes of the differential conductance are very close to each other in
the whole range of the bias voltage. In particular, both models lead to the
identical prediction: at the Kondo voltage, $eV=k\widetilde{T}_\text{K}$, the
differential conductance is equal to $2/3$ of its maximum.}
\end{figure}

A simple microscopic theory for $E_\alpha=E_\alpha^\text{R}+iE_\alpha^\text{I}$ is
obtained using the fact that in the deep Kondo regime (which we estimate as
$\mu_0-\epsilon_\text{d}\geqslant 4\Gamma$) in equilibrium and at
zero temperature the TDOS has maximum at $\epsilon\approx\mu_0$ and this maximum
is approximately equal to the unitary limit $2/\pi\Gamma$. This leads to the two
equations, $[\ln(2|E|/kT_\text{K})]\sin\phi=(\pi/2-\phi)\cos\phi$,
$[\ln(2|E|/kT_\text{K})]^2=\pi^2/4-(\pi/2-\phi)^2$, where
$kT_\text{K}\equiv 2W\exp[-\pi(\mu_0-\epsilon_\text{d})/\Gamma]$ is the Kondo
temperature, $|E|\equiv\sqrt{(E_\alpha^\text{R})^2+(E_\alpha^\text{I})^2}$ and
$\phi\equiv\arctan(E_\alpha^\text{I}/E_\alpha^\text{R})$. These equations show
that $\mathcal{E}_\alpha\equiv E_\alpha/kT_\text{K}$ is a universal ratio which
guarantees the correct scaling of the differential conductance. One gets
$\mathcal{E}_\alpha^\text{R}\approx 1.42$ and
$\mathcal{E}_\alpha^\text{I}\approx 1.05$. The universal ratio
$\mathcal{E}_\alpha^\text{I}/\mathcal{E}_\alpha^\text{R}$ provides the universal
phase, $\phi\approx 2/\pi$. Using this model of $\mathcal{E}_\alpha$ we can
calculate the differential conductance at any $T$ and $V$.

In Fig. \ref{figure_1} we compare the equilibrium results of our theory and NRG
for the highly asymmetric spin-$\frac{1}{2}$ SIAM. In our theory $U=\infty$. The
NRG results have been obtained for $U=250\Gamma$. Increasing $U$ further in NRG
does not produce any significant change in the universal characteristics. In both
models $\mu_0-\epsilon_\text{d}=8\Gamma$. The presence of the universal function in
the QD TDOS leads to the universal temperature dependence of the differential
conductance maximum (linear conductance) in the whole temperature range. As one
can see from the figure, the universal shape of the linear conductance obtained in
our theory is very close to the one in NRG.

Moreover, the differential conductance as a function of the bias voltage turns out to
be also universal. In Fig. \ref{figure_2} we show the nonequilibrium results of
our theory for the same model as in Fig. \ref{figure_1} and the RTRG results
\cite{Pletyukhov_2012}. One sees that the nonequilibrium universal shape of the
differential conductance for the asymmetric spin-$\frac{1}{2}$ SIAM with $U=\infty$ is
very close to the one in the $s-d$ model. In particular, in the deep nonequilibrium
crossover both theories predict the same result for the value of the differential
conductance at the Kondo voltage, namely, $2/3$ of its unitary limit value. This
universal result has been used \cite{Kretinin_2012} as a new experimental tool to
measure $T_\text{K}$.

At low energies we obtain \cite{supp_mat} the differential conductance
analytically,
$G(T,V)=(2e^2/h)[1-c_T(T/T_\text{K})^2-c_V(eV/kT_\text{K})^2]$. The absence of
linear terms in $G(T,V)$ proves that the low energy sector of our theory is
the Fermi liquid. It is known \cite{Oguri_2005,Sela_2009} that for the
symmetric SIAM $c_V/c_T=3/(2\pi^2)$. However, for the highly asymmetric
spin-$\frac{1}{2}$ SIAM with $U=\infty$ and finite $\mu_0-\epsilon_\text{d}$ little is
known on $c_V/c_T$. For this asymmetric model our theory predicts
\begin{equation}
c_T=\frac{4}{3}\,\frac{2\ln(2|\mathcal{E}_\alpha|)+1}{|\mathcal{E}_\alpha|^2},\quad\frac{c_V}{c_T}=c_a\frac{3}{2\pi^2},
\end{equation}
where $c_a=[4\ln(2|\mathcal{E}_\alpha|)+1]/[4\ln(2|\mathcal{E}_\alpha|)+2]\approx 0.86$.

At $T>T_\text{K}$ the linear conductance is well described by the analytical
expression \cite{supp_mat} obtained by Hamann \cite{Hamann_1967}.  The
Hamann's theory, based on the $s-d$ model, neglects charge fluctuations
present in the SIAM. Thus it underestimates the conductance
\cite{Costi_1994}. Our theory agrees well (inset in Fig. \ref{figure_1}) with
the Hamann's result until the asymmetry and charge fluctuations become
important at high $T$.

The results above show that the nonequilibrium universal behaviors of the highly
asymmetric spin-$\frac{1}{2}$ SIAM with $U=\infty$ and symmetric SIAM or $s-d$
model are close to each other. The difference between them partly results from
their fundamental physical difference ({\it e.g.}, the Fermi liquid coefficients,
like $c_V$, are sensitive to the asymmetry of the SIAM; see Ref. \onlinecite{Aligia_2012})
and partly from the quality of our theory and the quality of RTRG. Within our theory
the last point is equivalent to the question whether the theory with constant
$\mathcal{E}_\alpha$ may be improved through a dependence of $\mathcal{E}_\alpha$ on
the voltage and temperature. In Ref. \onlinecite{supp_mat} we show that a very simple
dependence of $\mathcal{E}_\alpha^\text{I}$ on the temperature and voltage may, indeed,
notably change the equilibrium and nonequilibrium results, obtained using the constant
model. Since the function $E_\alpha=(kT_\text{K})\mathcal{E}_\alpha$ effectively takes into
account higher order terms in $\Gamma$, a microscopic model for the temperature and
voltage dependence of $\mathcal{E}_\alpha$ may result from a proper renormalization group
theory. In particular, taking into account the quartic terms in the Keldysh effective
action and performing the 1- and 2-loop analysis may provide a renormalization of the
quadratic term and, as a result, a function $\mathcal{E}_\alpha=f(T/T_\text{K},eV/kT_\text{K})$.

In conclusion, we have proven the existence of a universal function $\mathcal{E}_\alpha$,
rigorously appearing in the formalism, and demonstrated how it determines the QD TDOS. A
simple microscopic model for $\mathcal{E}_\alpha$ has been proposed. This has allowed us
to unify the strong coupling regime, weak coupling regime and crossover region of the Kondo
state in equilibrium and nonequilibrium within a single analytical expression for the TDOS.
To demonstrate the practical importance of our theory we have applied it to the highly
asymmetric spin-$\frac{1}{2}$ SIAM with $U=\infty$. Our theory has provided the differential
conductance in the whole range of temperatures and voltages with the correct scaling
$T_\text{K}$, its Fermi liquid behavior at low energies, an analytical expression for the
ratio between the Fermi liquid coefficients, the prediction that at the Kondo voltage the
differential conductance is equal to $2/3$ of its maximum, an excellent agreement with known
theories at high energies. At the same time it has raised a challenge to develop a
renormalization group method to improve the quality of the present theory.

The authors thank Mikhail Pletyukhov and Herbert Schoeller for providing the RTRG data,
Theodoulos Costi and Lukas Merker for providing the NRG data and all of them for fruitful
discussions.

Support from the DFG under the program SFB 689 is acknowledged.

\end{document}